\documentclass[12pt]{iopart}

\usepackage{iopams}
\usepackage{graphicx}

\begin{document}


\title{Stable propagation of a modulated positron beam in a bent crystal channel}

\author{A.~Kostyuk$^{1}$, A.V.~Korol$^{1,2}$, A.V.~Solov'yov$^{1}$ and 
W. Greiner$^{1}$}



\address{
$^{1}$ Frankfurt Institute for Advanced Studies,
 Johann Wolfgang Goethe-Universit\"at, 
Ruth-Moufang-Str.~1, 60438 Frankfurt am Main, Germany\\
$^{2}$ Department of Physics,
St Petersburg State Maritime Technical University,
St Petersburg, Russia
}



\begin{abstract}
The propagation of a modulated positron beam  in a planar crystal channel
is investigated. 
It is demonstrated that the beam preserves its modulation at sufficiently large
penetration depths which opens the prospect
of using a crystalline undulator as a coherent source of hard x-rays.
This finding is a crucial milestone in developing a new type of lasers radiating
in the hard x-ray and gamma-ray range.
\end{abstract}




\submitto{\JPB}


\maketitle

In this communication we study for the first time the evolution of a modulated 
particle beam in a bent planar crystal channel and demonstrate that 
a positron beam preserves 
its modulation at sufficiently large
penetration depths,  which opens the prospect
of using a crystalline undulator as a coherent source of hard x-rays.
Solving this problem is of crucial 
importance in the theory of the crystal undulator based laser (CUL) \cite{first,KSG1999,klystron}
--- a new electromagnetic radiation source in hard x- and gamma-ray range.

Channelling takes place if charged particles enter a single crystal at small
angle with respect to crystallographic planes or axes \cite{Lindhard}. The 
particles get confined
by the interplanar or axial potential and follow the shape of the corresponding 
planes and axes. This suggested the idea \cite{Tsyganov1976} of using bent crystals
to steer the particles beams. Since its first experimental verification \cite{Elishev1979}
the idea to deflect or extract  high-energy charged particle beams by means of tiny bent crystals
replacing huge dipole magnets has been attracting a lot of interest worldwide. Bent crystal
have been routinely used for beam extraction in the Institute for High Energy Physics, Russia
\cite{Afonin2005}. A series of experiments on the bent crystal deflection of proton and heavy ion beams was
performed at different
accelerators \cite{Arduini1997,Scandale2008,Carrigan1999,Fliller2006,Strokov2007} throughout the world.
The bent crystal method has been proposed to extract particles from the beam halo at
CERN Large Hadron Collider \cite{Uggerhoj2005}
The possibility of deflecting positron \cite{Bellucci2006} and electron \cite{Strokov2007,Strokov2006}
beams has been studied as well.

\begin{figure}[ht]
\begin{center}
\includegraphics*[width=15cm]{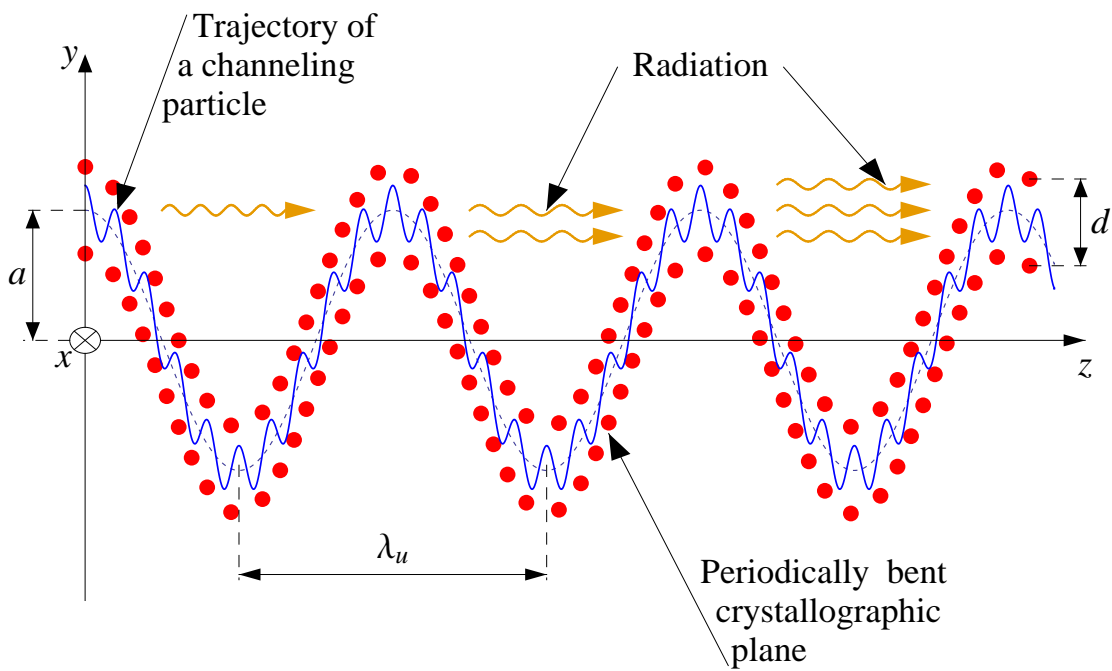}
\end{center}
\caption{Schematic representation of the crystalline undulator.
}
\label{undulator.fig}
\end{figure}

A single crystal with {\it periodically} bent crystallographic planes can force 
channelling particles to move along  nearly sinusoidal trajectories
and radiate in the hard x- and gamma-ray frequency range (see figure \ref{undulator.fig}).
The feasibility of such a device, known as the 'crystalline undulator`,
was demonstrated theoretically a decade ago \cite{first} (further
developments as well as historical references are reviewed in \cite{KSG2004_review}).
More recently, an electron based crystalline undulator has been proposed 
\cite{Tabrizi}. 

It was initially proposed to obtain sinusoidal bending
by the propagation of an acoustic wave along the crystal 
\cite{first,KSG1999}. The advantage of this approach is its flexibility:
the period of deformation can be chosen by tuning the frequency of the ultrasound.
However, this approach is rather challenging technologically and yet to be tested 
experimentally.
Several other technologies for the  manufacturing of periodically bent crystals
have been developed and tested. These include
making regularly spaced grooves on the crystal surface
either by a diamond blade \cite{BellucciEtal2003,GuidiEtAl_2005}
or by means of laser-ablation \cite{Balling2009},
deposition
of periodic
Si$_3$N$_4$ layers onto the surface of a Si crystal \cite{GuidiEtAl_2005},
growing of Si$_{1-x}$Ge$_x$ crystals\cite{Breese97}
with a periodically varying Ge content $x$ \cite{MikkelsenUggerhoj2000,Darmstadt01}.

Experimental studies of the crystalline undulator are currently in progress. The first results 
are reported in \cite{Baranov2006} and \cite{Backe2008}.


The advantage of the crystalline undulator is in extremely strong
electrostatic fields inside a crystal which are able
to steer the particles much more effectively than even the most advanced
superconductive magnets. 
This fact allows to make the period $\lambda_\mathrm{u}$ of the crystalline undulator
in the range of hundreds or tens micron which is two to three orders of 
magnitude smaller than that of conventional undulator. Therefore
the wavelength of the produced radiation 
$\lambda \sim \lambda_\mathrm{u}/(2 \gamma^2)$ ($\gamma \sim 10^3$--$10^4$ being the 
Lorentz factor of the particle) can reach the (sub)picometer range, 
where conventional sources with 
comparable intensity are unavailable \cite{Topics}.

Even more powerful and coherent radiation will be emitted if
the probability density of the particles in the beam is modulated
in the longitudinal direction with the period $\lambda$, equal
to the wavelength of the emitted radiation.
In this case, the electromagnetic waves emitted  in the forward direction by
different particles have approximately the same phase \cite{Ginzburg}. Therefore,
the intensity of the radiation becomes proportional
to the beam density squared (in contrast to the linear proportionality for an unmodulated
beam). This increases the photon flux {\it by orders of magnitude}
relative to the radiation of unmodulated beam of the same density.
The radiation of a modulated beam in an undulator is 
a keystone of the physics of free-electron lasers (FEL) \cite{Madey,SchmueserBook}.
It can be considered as  a
classical counterpart of the stimulated  emission in quantum physics.
Therefore, if similar phenomenon takes place in a crystalline undulator, it can be
referred to as the {\it lasing regime of the crystalline undulator}.


The feasibility of CUL radiating in the hard
x-ray and gamma-ray range was considered for the fist time in \cite{first,KSG1999}. 
Recently,
a two-crystal scheme, the gamma klystron, has been proposed \cite{klystron}.

\begin{figure}[ht]
\begin{center}
\includegraphics[width=15cm]{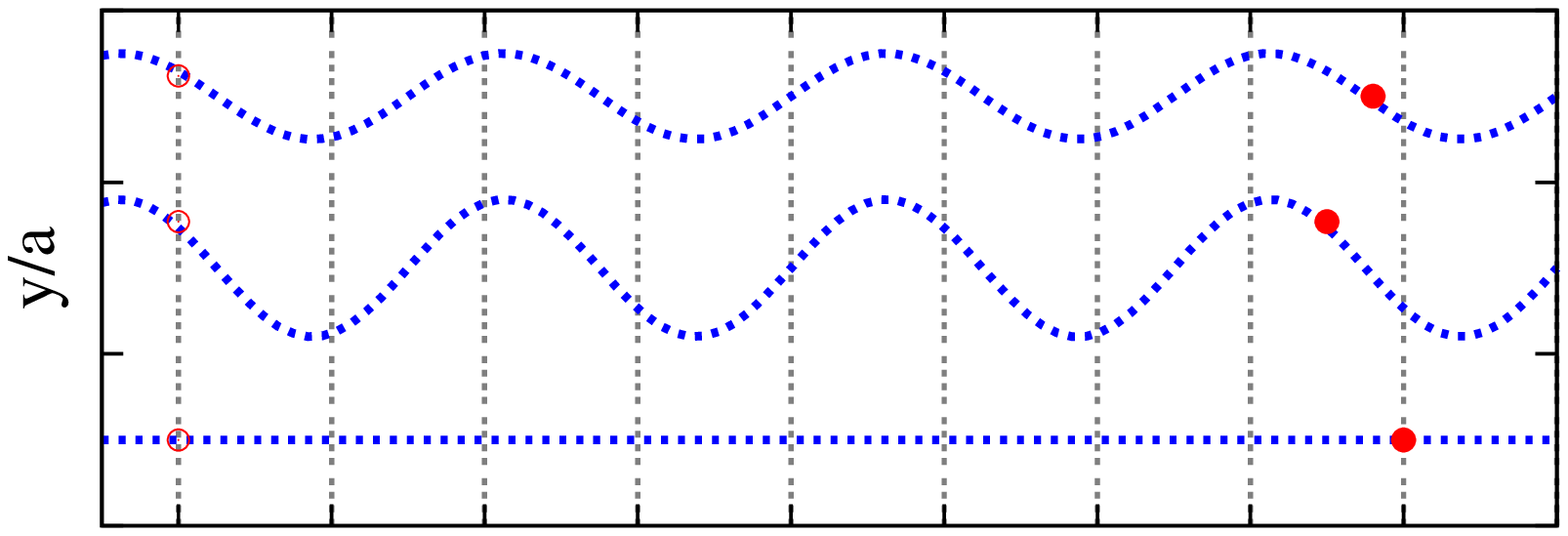}
\includegraphics[width=15cm]{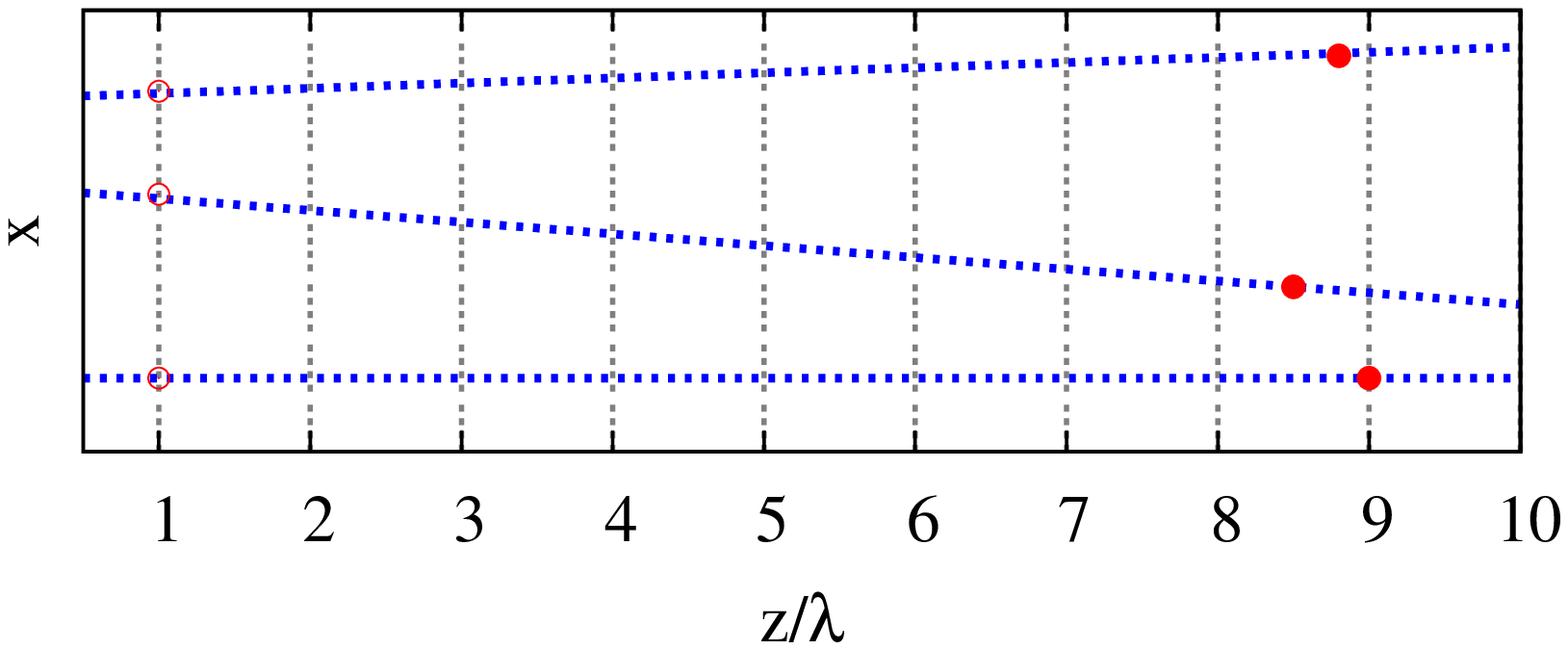}
\end{center}
\caption{Due to different amplitudes of channelling oscillation
(upper panel) and different momentum directions 
in the $(xz)$ plane (lower panel), the initially modulated
beam gets demodulated. The open and filled circles denote the
same particles at the crystal entrance and after
travelling some distance in the crystal channel, respectively.
}
\label{demodulation.fig}
\end{figure}

A simplified model used in the cited papers assumed that all particle trajectories
follow exactly the shape of the bent channel. In reality, however, the particle
moving along the channel also oscillates in the transverse direction
with respect 
to the channel axis (see the shape of the trajectory 
in figure \ref{undulator.fig}). Different particles have
different amplitudes of the oscillations inside the channel
(figure \ref{demodulation.fig}, upper panel).
Similarly, the directions of particle momenta
in $(xz)$ plane are slightly different
(figure \ref{demodulation.fig}, lower panel).
Even if the speed of the particles along their
trajectories is the same, the particles oscillating with 
different amplitudes or the particles with different trajectory
slopes with respect to $z$ axis have slightly
different components of their velocities along the channel.
As a result, the beam gets demodulated.
An additional contribution to the beam demodulation comes 
from incoherent collisions
of the channelling particles with the crystal constituents.

In the case of an unmodulated beam, 
the length of the crystalline undulator 
and, consequently, the maximum accessible intensity of the radiation
is limited by the dechannelling process.
The channelling particle gradually gains the energy of transverse 
oscillation due to collisions with crystal constituents.
At some point this energy exceeds the maximum value of the interplanar
potential and the particle leaves the channel. The average penetration length 
at which this happens is known as the {\it dechannelling length}.
The dechannelled particle no longer follows the sinusoidal 
shape of the channel and, therefore, does not contribute to the 
undulator radiation. Hence, 
the reasonable length of the crystalline undulator 
is limited to a few dechannelling lengths. A longer crystal would attenuate rather
then produce the radiation. Since the intensity of the undulator radiation is
proportional to the undulator length squared, the dechannelling length 
and the attenuation length are the 
main restricting factors that have to be taken into account when the radiation output 
is calculated.

In contrast, not only the shape of the trajectory but also the particles positions 
with respect to each other along $z$ axis are important for the lasing regime.
If these positions become random because of the beam demodulation, the intensity
of the radiation drops even if the particles are still in the channelling mode.
Hence, it is the beam demodulation rather than dechannelling that restricts the 
intensity of the radiation of CUL.
Understanding this process and estimating the characteristic length at which this phenomenon
takes place is, therefore, a cornerstone of the theory of this
new radiation source.

Let us consider the distribution 
$f(t,s;\xi,E_{y})$
of the beam particles 
with respect to the angle between the particle trajectory
and the $z$ axis in the $(xz)$ plane 
$\xi = \arcsin p_{x}/p \approx p_{x}/p$ and the energy of the channelling
oscillation $E_{y} = p_{y}^2/2 E + U(y)$ \footnote
{We chose the system of units in such 
a way that the speed of light is equal to unity.
Therefore, mass, energy and momentum have the same dimensionality. This is also true
for length and time.}. Here $p$, $p_{x}$ and $p_{y}$ are, respectively, the particle momentum and its $x$ and $y$ components,
$U(y)$ is the interplanar potential which in the case of positive particles, which channel between the 
crystallographic planes, can be approximated by a parabola $U(y)=  U_{\max} (y/y_{\max})^2$, and 
$E$ is the particle energy (we will consider only ultrarelativistic 
particles, therefore $E \approx p$).
It can be shown that the
evolution of this distribution in the  crystal
channel with the time $t$ and the
longitudinal coordinate (penetration depth
into the crystal along the curved channel) $s$ can be described by the following differential
equation of Fokker-Planck type:
\begin{equation}
\frac{\partial f}{\partial t} +  
\frac{\partial f}{\partial s} v_{s}
 = D_0 \left [ \frac{\partial }{\partial E_{y}}
\left(E_{y}
\frac{\partial f }{\partial E_{y}}
\right)
+
\frac{1}{E} 
\frac{\partial^2 f }{\partial \xi^2}
\right ] \ .
\label{diffeqD_0}
\end{equation} 
Here $D_0$ is the diffusion coefficient that is dominated 
by the scattering of the beam particles by lattice electrons.
The particle longitudinal velocity
averaged over the undulator period, $v_{s}$,  is given by
\begin{equation}
v_{s} =    \left( 
1 - \frac{1}{2 \gamma^2} - \frac{\xi^2}{2} - \frac{E_{y}}{2 E}
\right).
\end{equation}
Equation (\ref{diffeqD_0}) is akin to the equation describing 
dechannelling process (see e.g. \cite{BiryukovChesnokovKotovBook})
and is derived in a similar way.
The novel feature of it is the presence of time variable, which allows us
to describe time dependent (modulated) beams. Additionally, it takes 
into account scattering in the $(x,z)$ plane.

If the beam is periodically modulated (bunched) the distribution
can be represented as a Fourier series:
\begin{equation}
f(t,s;\xi,E_{y}) = \sum_{j=-\infty}^{\infty} g_{j} (s;\xi,E_{y}) \exp (i j \omega t),
\end{equation}
with $g_{j}^{*} (s;\xi,E_{y}) = g_{-j} (s;\xi,E_{y})$ to ensure the real value of the
particle distribution. Since Eq. (\ref{diffeqD_0}) is linear, 
it is sufficient to consider only one harmonic.
Substituting 
$f(t,s;\xi,E_{y}) = g (s;\xi,E_{y}) \exp (i \omega t)$ one obtains
\begin{equation}
i \omega g(s;\xi,E_{y}) +
 \frac{\partial g}{\partial s} 
v_{s}
 = D_0 \left [ \frac{\partial }{\partial E_{y}}
\left(E_{y}
\frac{\partial g }{\partial E_{y}}
\right)
+
\frac{1}{E} 
\frac{\partial^2 g }{\partial \xi^2}
\right ] .
\label{diffeqg}
\end{equation}

To simplify this equation, we make the substitution
$$g(s;\xi,E_{y}) = \exp \left( - i \omega s \right)
\tilde{g}(s;\xi,E_{y})$$
and assume that the variation of  
$\tilde{g}(s;\xi,E_{y})$ within the modulation period is small:
$\partial \tilde{g} / \partial s \ll \omega \tilde{g}(s;\xi,E_{y})$.
This allows us to neglect the terms 
$(1-v_{s}) \partial \tilde{g} / \partial s$
while keeping the terms 
$(1-v_{s}) \omega \tilde{g}(s;\xi,E_{y})$.
The resultant partial differential equation
for $\tilde{g}(s;\xi,E_{y})$
can be solved by the method of separation of 
variables. Putting
$\tilde{g}(s;\xi,E_{y}) = \mathcal{S}(s) \Xi(\xi) \mathcal{E} (E_{y})$,
we obtain a set of ordinary differential equations:
\begin{eqnarray}
\frac{D_0}{E} 
\frac{1}{\Xi(\xi)} 
\frac{d^2 \Xi(\xi)}{d \xi^2} 
- i \omega \frac{\xi^2}{2}
&=& \mathcal{C}_{\xi} , \label{eqXi} \\
\frac{D_0}{\mathcal{E}(E_{y})}
\frac{d }{d E_{y}}
\left(E_{y}
\frac{d  \mathcal{E}(E_{y})}{d E_{y}}
\right) 
- i \omega \frac{E_{y}}{2 E}
&=&  \mathcal{C}_{y}, \label{eqE} \\
\frac{1}{\mathcal{S}(s)}
\frac{d \mathcal{S}(s)}{d s} + 
\frac{i \omega}{2 \gamma^2} &=& \mathcal{C}_{s} , \label{eqS} 
\end{eqnarray}
where $\mathcal{C}_{s}$, $\mathcal{C}_{\xi}$ and $\mathcal{C}_{y}$ do not
depend on any of the variables $s$, $\xi$ and $E_{y}$ and satisfy the condition 
\begin{equation}
\mathcal{C}_{s} = \mathcal{C}_{\xi} + \mathcal{C}_{y} .
\label{sumC}
\end{equation}

Eq. (\ref{eqXi}) has the form of the Schr{\"o}dinger 
equation for the harmonic oscillator.
Its eigenvalues and eigenfunctions are, respectively,
\begin{equation}
\mathcal{C}_{\xi,n}
 = - (1+i)
\sqrt{\frac{\omega D_0}{E}} \left( n + \frac{1}{2} \right),
\ \ \ n = 0,1,2, \dots
\label{Cxi}
\end{equation}
and
\begin{equation}
\Xi_{n}(\xi) =
H_{n} \left( \mathrm{e}^{i \pi / 8} \sqrt[4]{\frac{\omega E}{2 D_0}} \; \xi \right) 
\exp \left( -  \frac{1 + i}{4} \sqrt{\frac{\omega E}{D_0}} 
\xi^2 \right).
\label{Xin}
\end{equation}
Here
$H_{n}(\dots)$ are Hermite polynomials.

Eq. (\ref{eqE}) can be reduced to the Laguerre differential equation, so that 
its solution can be represented as
\begin{equation}
\mathcal{E}_{k}(E_{y})
\! = \!
\exp \!
\left( \!
- \frac{1+i}{2} \sqrt{\frac{\omega}{D_{0} E}} E_{y}
\right)
L_{\nu_{k}} \! \!
\left( \!
(1+i) \sqrt{\frac{\omega}{D_{0} E}} E_{y} \!
\right)
\label{Ek}
\end{equation}
where $L_{\nu}(\dots)$ is the Laguerre function\footnote{At nonnegative integer
values of $\nu$, the Laguerre function is reduced to the well known Laguerre polynomials.
In the general case that is relevant to our consideration, it can be represented by 
an infinite series:
$L_{\nu}(\varepsilon) =  \sum_{j=0}^{\infty} \prod_{m=0}^{j-1} (m - \nu) 
\varepsilon^j / (j!)^2$.} and $\nu_{k}$  is related to the eigenvalue  
$\mathcal{C}_{y,k}$ via
\begin{equation}
\mathcal{C}_{y,k} = 
- \frac{(1+i)}{2} 
\sqrt{\frac{D_{0} \, \omega}{E}} (2 \nu_{k} + 1),
\ \ \ k=1,2,3,\dots
\label{Cynu}
\end{equation}
The eigenvalues can be found by imposing the boundary conditions.
The maximum energy  of channelling oscillations in a bent channel
with parabolic potential
is given by $E_{y}^{\max} = U_{\max} (1-C)^2$, where
$C = F_\mathrm{c}/U_{\max}'$ is defined as a
ratio of the centrifugal force $F_\mathrm{c}$ to 
the maximum value of the interplanar force $U_{\max}'$ \cite{KSG2004_review}.
Therefore, the density boundary condition has the form:
\begin{equation}
L_{\nu_{k}} 
\left( 
(1+i) \sqrt{\frac{\omega}{D_{0} E}} U_{\max} (1-C)^2
\right) = 0.
\label{boundcond}
\end{equation}
Equation (\ref{boundcond}) has to be solved  for $\nu_{k}$.
(the subscript $k$ enumerates different roots of the 
equation) and the result has to be substituted into (\ref{Cynu}).

It is convenient to represent the eigenvalues in the form
\begin{equation}
\mathcal{C}_{y,k} = - \frac{\alpha_k(\kappa, C)}{L_\mathrm{d}}  
- i \omega \theta_\mathrm{L}^2 \beta_k(\kappa, C).
\end{equation}
Here $L_\mathrm{d} = 4 U_{\max} /(j_{0,1}^2 D_0)$ is the 
dechannelling length in a straight channel 
\cite{BiryukovChesnokovKotovBook}
($j_{0,k}$  is $k$-th zero of the Bessel function $J_0(\varepsilon)$), 
$\theta_\mathrm{L} = \sqrt{2 U_{\max}/E}$ is the corresponding 
Lindhard's angle. We introduce the parameter
\begin{equation}
\kappa = \pi \frac{L_\mathrm{d}}{\lambda}  \theta_\mathrm{L}^2,
\label{kappa}
\end{equation}
where $\lambda = 2 \pi / \omega$ is the spatial period of the modulation.
The functions $\alpha_k(\kappa, C)$ and $\beta_k(\kappa, C)$
can be represented as
\begin{eqnarray}
\alpha_k(\kappa, C) & = & \frac{\alpha_k \left( \kappa (1-C)^4 \right)}{(1-C)^2} \\
\beta_k(\kappa, C)  & = & (1-C)^2 \beta_k \left( \kappa (1-C)^4 \right)
\end{eqnarray}
Here $\alpha_k(\kappa) \equiv \alpha_k(\kappa, C=0)$ and $\beta_k(\kappa) \equiv \beta_k(\kappa, C=0)$
are the cooresponding functions for the straight channel which are found
by solving numerically Eq. (\ref{boundcond}) combined with
(\ref{Cynu}). 

Using (\ref{sumC}), one finds the solution 
of Eq. (\ref{eqS}):
\begin{eqnarray}
\mathcal{S}_{n,k}(s) &=& \exp \left \{
- \frac{s}{L_\mathrm{d}} 
\left[ \alpha_k(\kappa, C) + (2 n + 1) 
\frac{\sqrt{\kappa}}{j_{0,1}}
\right ]
\right. 
\label{Snk}
\\
& & 
\left.
-
i \omega s
\left [
\frac{1}{2 \gamma^2} + 
\theta_\mathrm{L}^2 \beta_k(\kappa, C) + 
\theta_\mathrm{L}^2  \frac{(2 n + 1)}{2 j_{0,1} \sqrt{\kappa}}
\right ]
\right \} .
\nonumber
\end{eqnarray}

Hence, the solution of Eq. (\ref{diffeqg}) is
represented as
\begin{equation}
g(s;\xi,E_{y}) = \exp \left( - i \omega s \right) \sum_{n=0}^{\infty} \sum_{k=1}^{\infty}
\mathfrak{c}_{n,k}
\Xi_{n}(\xi) \mathcal{E}_{k}(E_{y}) \mathcal{S}_{n,k}(s) ,
\end{equation}
where the coefficients $\mathfrak{c}_{n,k}$ are found from the particle distribution at the
entrance
of the crystal channel.
Due to the exponential decrease of $\mathcal{S}_{n,k}(s)$ with $s$
(see (\ref{Snk})), the asymptotic behaviour of $\tilde{g}(s;\xi,E_{y})$ 
at large $s$
is dominated by the term with $n=0$ and $k=1$
having the smallest value of the
factor $\left[ \alpha_k(\kappa, C) + (2 n + 1) \sqrt{\kappa}/j_{0,1}
\right ]$ in the exponential.
Therefore, at sufficiently large penetration depths, the particle
distribution depends on $s$ as
$
g (s;\xi,E_{y}) \propto \exp \left (
- s / L_{\mathrm{dm}}  - i \omega/u_{s} \, s
\right) 
$
where $ L_{\mathrm{dm}}$ is the newly introduced parameter --- {\it the demodulation 
length}:
\begin{equation}
L_{\mathrm{dm}} = \frac{ L_\mathrm{d} }{\alpha_1(\kappa, C) + \sqrt{\kappa}/j_{0,1}}
\label{Ldm}
\end{equation}
and $u_{s}$ is the phase velocity of the modulated beam along the crystal channel
\begin{equation}
u_{s} = \left [ 1 + \frac{1}{2 \gamma^2} + 
\theta_\mathrm{L}^2 \left( \beta_k(\kappa, C) + \frac{1}{2 j_{0,1} \sqrt{\kappa}} \right) 
 \right  ]^{-1} .
\label{us}
\end{equation}
This parameter is important for establishing the resonance conditions between 
the undulator parameters and the radiation wavelength. It
will be analysed elsewhere.

\begin{figure}[ht]
\begin{center}
\includegraphics*[width=15cm]{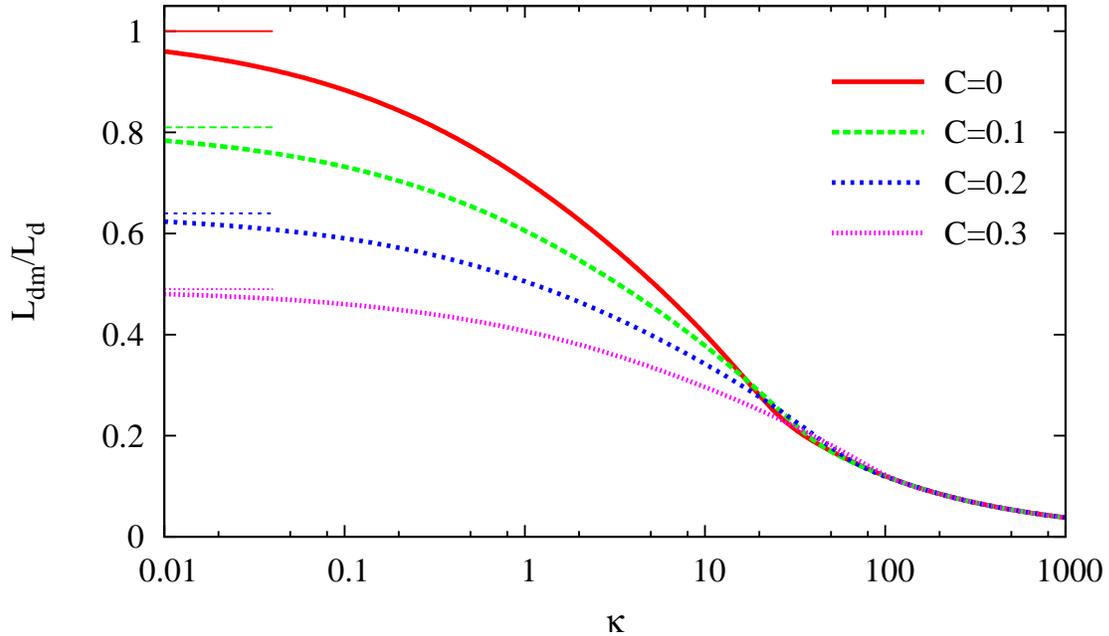}
\end{center}
\caption{The ratio of the demodulation length 
$L_{\mathrm{dm}}$ (\ref{Ldm})
to the dechannelling length in the straight channel $L_\mathrm{d}$ versus 
the parameter $\kappa$ (\ref{kappa}) for different values of factor $C$.
The corresponding asymptotic values at $\kappa \rightarrow 0$ 
are shown by thin horizontal lines.}
\label{Ldm.fig}
\end{figure}

In this communication we concentrate our attention on the demodulation length.
It represents the characteristic scale of the penetration depth at
which an initially modulated beam of channelling particles becomes demodulated.

Figure \ref{Ldm.fig} presents the dependence of the ratio $L_{\mathrm{dm}}/L_\mathrm{d}$ on
the parameter $\kappa$. At $\kappa \rightarrow 0$, the demodulation length approaches 
$(1-C)^2 L_\mathrm{d}$ which is the dechannelling length in the bent crystal.
It was proven for a number
crystals channels \cite{KSG2004_review}
that the dechannelling length of positrons  is sufficiently large to make the crystalline undulator feasible.
Such a crystalline undulator becomes a CUL, i.e. it generates coherent radiation, provided that it is
fed by a modulated positron beam and the beam preserves its modulation over the length of the crystal.
This takes place if the demodulation length in the crystalline undulator is not much smaller
than the dechannelling length. As is seen from the figure, the demodulation 
length is smaller than dechannelling length by only 20--30\% at $\kappa \lesssim 1$ for $C$ varying between
$0$ and  $0.3$. It noticeably drops, however, at $\kappa \gtrsim 10$.
Hence, CUL is feasible if there exist crystal channels ensuering $\kappa \lesssim 1$
in the range of the photon energies above $\sim 100$ keV.\footnote{X-rays with photon energies of  a few 
tens keV or less are strongly absorbed
in the crystal. This puts the lower limit on the energies of the photons that can be generated by
crystalline undulator based devices.}

\begin{figure}[ht]
\begin{center}
\includegraphics*[width=15cm]{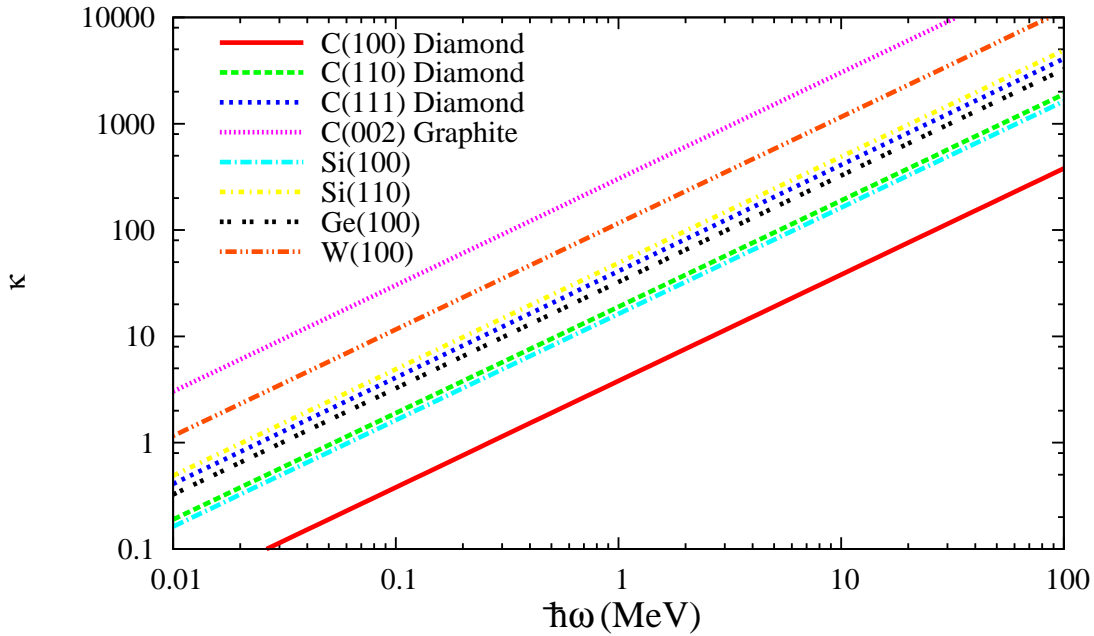}
\end{center}
\caption{The parameter $\kappa$ (\ref{kappa}) versus the photon energy
$\hbar \omega$ for different crystals and crystallographic 
planes.}
\label{hbaromega_kappa.fig}
\end{figure}

Indeed, such crystal channels do exist. Figure \ref{hbaromega_kappa.fig} shows the dependence of 
the parameter $\kappa$ on the energy of the emitted photons $\hbar \omega = 2 \pi \hbar/\lambda$ for
different crystal channels. The calculation was done for 1 GeV\footnote{Note that $\kappa$
depends weakly (logarithmically) on the particle energy. Therefore,
changing the beam energy by an order of magnitude would leave
figure \ref{hbaromega_kappa.fig} practically unchanged.} positrons using the formula 
for the dechannelling length from \cite{KSG2004_review,BiryukovChesnokovKotovBook}.
As one sees from the figure,  $\kappa \sim 1$ corresponds to $\hbar \omega = 100-300$ keV
for (100) and (110) planes in Diamond and  (100) plane in Silicon. So these channels are 
the most suitable candidates for using in CUL. This is, however, not the case for a number of other crystals
e.g. for graphite and tungsten having $\kappa \gtrsim 10$ in the same photon energy range.

At $\hbar \omega \sim 10$ MeV, $\kappa$ becomes larger than $10$ for all crystal channels.
This puts the upper limit on the energies of the photons that can be generated by CUL.
It is expected to be most successful in the hundred keV range, while generating MeV photons
looks more challenging.

According to our estimations, the brilliance as high as\linebreak
$10^{25}$--$10^{26}$ 
$\mbox{Photons}/(\mbox{s} \, \mbox{mm}^2 \,  \mbox{mrad}^2 \,  0.1\% \mbox{BW})$ can be obtained
in a CUL fed by a completely modulated positron 
beam with current $1$ kA and particle density $10^{18}$ cm$^{-3}$.

One may expect that the demodulation is not limited to the processes illustrated in
figure \ref{demodulation.fig}. An additional contribution can come from the energy spread of the 
channelling particles, as it usually happens in ordinary FELs.
In fact, the contribution of the energy spread to the beam demodulation on the distance of
a few dechannelling lengths is negligible. It would be substantial if the relative 
{\it spread} $\delta E / E$ of particle energies
would be comparable to or larger than the ratio $\lambda_\mathrm{u}/L_{d}$. The latter 
ratio, however, can not be made smaller than $10^{-2}$ \cite{KSG2004_review},\footnote{Note that the 
corresponding quantity in ordinary ultraviolet and soft x-ray FELs, the inverse number of undulator periods 
$1/N_\mathrm{u} = \lambda_\mathrm{u}/L$, is usually of the order of $10^{-3}$--$10^{-4}$ \cite{SchmueserBook}.
That is why these FELs are so demanding to the small energy spread of the electron beam.}
while modern accelerators usually have a much smaller relative energy spread. The same is true for the energy spread
induced by the stochastic energy losses of the channelling particles due to the interaction 
with the crystal constituents and the radiation of photon. It was shown in Ref.\cite{KSG2000} that at initial
energies of $\sim 1 GeV$ or smaller, the average relative energy {\it losses} of a positron in the crystalline undulator 
$\Delta E / E$ are smaller than $10^{-2}$. Clearly, the induced energy {\it spread} $\delta E / E \ll \Delta E / E$
is safely below the ratio $\lambda_\mathrm{u}/L_{d}$.
From these reasons, we ignored energy spread of the particles in our calculations.


In conclusion, we have studied the propagation of a modulated positron beam in a bent planar
crystal channel.
It has been demonstrated that one can find the crystal channels in which the beam preserves 
its modulation at the penetration depths  sufficient for producing coherent radiation
with the photon energy of hundreds of keV. This opens the prospects for creating intense monochromatic 
radiation sources in a frequency range which is unattainable for conventional free electron lasers.
Developing suitable methods of beam modulation would be the next milestone on the way towards this goal.


\mbox{}\\

This work has been supported in part by the European Commission 
(the PECU project, Contract No. 4916 (NEST)) and by
Deutsche Forschungsgemeinschaft.

\newpage

\end{document}